\documentclass[floatfix,aps,prl,preprint,superscriptaddress]{revtex4}
\usepackage{graphicx}
\usepackage[]{epsfig}
\usepackage{times,amsmath,amssymb}

\usepackage{color}

\begin{document}

\title{Channel length dependence of the formation of quantum dots in GaN/AlGaN FETs}

\author{Kazuma Matsumura}
\affiliation{Research Institute of Electrical Communication, Tohoku University, 2-1-1 Katahira, Aoba-ku, Sendai 980-8577, Japan}
\affiliation{Graduate School of Engineering, Tohoku University, 6-6 Aramaki Aza Aoba, Aoba-ku, Sendai 980-0845, Japan}

\author{Takaya Abe}
\affiliation{Research Institute of Electrical Communication, Tohoku University, 2-1-1 Katahira, Aoba-ku, Sendai 980-8577, Japan}
\affiliation{Graduate School of Engineering, Tohoku University, 6-6 Aramaki Aza Aoba, Aoba-ku, Sendai 980-0845, Japan}

\author{Takahito Kitada}
\affiliation{Research Institute of Electrical Communication, Tohoku University, 2-1-1 Katahira, Aoba-ku, Sendai 980-8577, Japan}
\affiliation{Graduate School of Engineering, Tohoku University, 6-6 Aramaki Aza Aoba, Aoba-ku, Sendai 980-0845, Japan}

\author{Takeshi Kumasaka}
\affiliation{Research Institute of Electrical Communication, Tohoku University, 2-1-1 Katahira, Aoba-ku, Sendai 980-8577, Japan}

\author{Norikazu Ito}
\affiliation{ROHM Co., Ltd, 21 Saiinnmizosakicho, Ukyo-ku, Kyoto, Kyoto 615-8585, Japan}

\author{Taketoshi Tanaka}
\affiliation{ROHM Co., Ltd, 21 Saiinnmizosakicho, Ukyo-ku, Kyoto, Kyoto 615-8585, Japan}

\author{Ken Nakahara}
\affiliation{ROHM Co., Ltd, 21 Saiinnmizosakicho, Ukyo-ku, Kyoto, Kyoto 615-8585, Japan}

\author{Tomohiro Otsuka}
\email[]{tomohiro.otsuka@tohoku.ac.jp}
\affiliation{WPI Advanced Institute for Materials Research, Tohoku University, 2-1-1 Katahira, Aoba-ku, Sendai 980-8577, Japan}
\affiliation{Research Institute of Electrical Communication, Tohoku University, 2-1-1 Katahira, Aoba-ku, Sendai 980-8577, Japan}
\affiliation{Graduate School of Engineering, Tohoku University, 6-6 Aramaki Aza Aoba, Aoba-ku, Sendai 980-0845, Japan}
\affiliation{Center for Science and Innovation in Spintronics, Tohoku University, 2-1-1 Katahira, Aoba-ku, Sendai 980-8577, Japan}
\affiliation{Center for Emergent Matter Science, RIKEN, 2-1 Hirosawa, Wako, Saitama 351-0198, Japan}

\date{\today}

\begin{abstract}
Quantum dots can be formed in simple GaN/AlGaN field-effect-transistors (FETs) by disordered potential induced by impurities and defects.
Here, we investigate the channel length dependence of the formation of quantum dots.
We observe decrease of the number of formed quantum dots with decrease of the FET channel length.
A few quantum dots are formed in the case with the gate length of 0.05~$\mu $m and we evaluate the dot parameters and the disordered potential.
We also investigate the effects of a thermal cycle and illumination of light, and reveal the change of the disordered potential.
\end{abstract}

\maketitle

Semiconductor quantum dots confine electron in small regions and the quantum nature of electrons appears.
The confinement potential can be formed artificially by etching or gating of semiconductor nanostructures.
From the viewpoint of fundamental physics, the inner electronic states in quantum dots have been investigated\cite{1996TaruchaPRL, 1997KouwenhovenScience, 2000CiorgaPRB, 2001KouwenhovenRepProgPhys} and physics in combined systems were explored\cite{1995YacobyPRL, 1998GoldhaberGordonNature, 2000vanderWielScience, 2002KobayashiPRL}.
Also, the quantum dots have applications in low-power devices including single-electron transistors\cite{1992KastnerRMP, 1996ChenAPL} and devices for quantum information processing\cite{1998LossPRA, 2010LaddNature}.
Especially for semiconductor spin qubits, readout\cite{2002OnoScience, 2004ElzermanNature} and control\cite{2005PettaScience, 2006KoppensNature, 2014YonedaPRL, 2014VeldhorstNatureNano, 2018YonedaNatureNano, 2021KojimaNpjQuantumInfo} of the spin quantum states have been demonstrated and intensively studied in GaAs and Si-based deices.

Quantum dots can also be formed in simple FET structures by intrinsic disordered potential induced by impurities and defects.
Formation of quantum dots in FET have been reported and high temperature qubit operation have been demonstrated\cite{2006SellierPRL, 2007OnoAPL, 2008LansbergenNatPhy, 2010TabePRL, 2019OnoSciRep}.
Such quantum transport in FET structures have been reported mainly in Si-based devices.
On the other hand, high-mobility transistors based on GaN/AlGaN heterostructures\cite{1999AmbacherJApplPhys, 2004ManfraAPL} will be another possible material.
New quantum devices can be expected by utilizing the wide and direct band gap in GaN which work at higher temperatures and couple with light.
Also, the quantum transport will be useful to understand the actural potential formed in GaN/AlGaN FETs and improve the performance.
Previously, we investigated the formation of multiple quantum dots in GaN/AlGaN FETs near the pinch-off condition by the disordered potential near the FET conduction channel\cite{2020OtsukaSciRep}. 

In this paper, we investigate the channel length dependence of the formation of quantum dots in GaN/AlGaN FETs.
By decreasing the channel length, there are fewer impurities and defects which contribute to quantum dot formation, and the number of quantum dots can be reduced.
A few quantum dots are formed in a device with the short gate length and we evaluate the dot parameters and the potential.
We also evaluate effects of a thermal cycle and illumination of light, and reveal change of the disordered potential.

\begin{figure}
\begin{center}
  \includegraphics{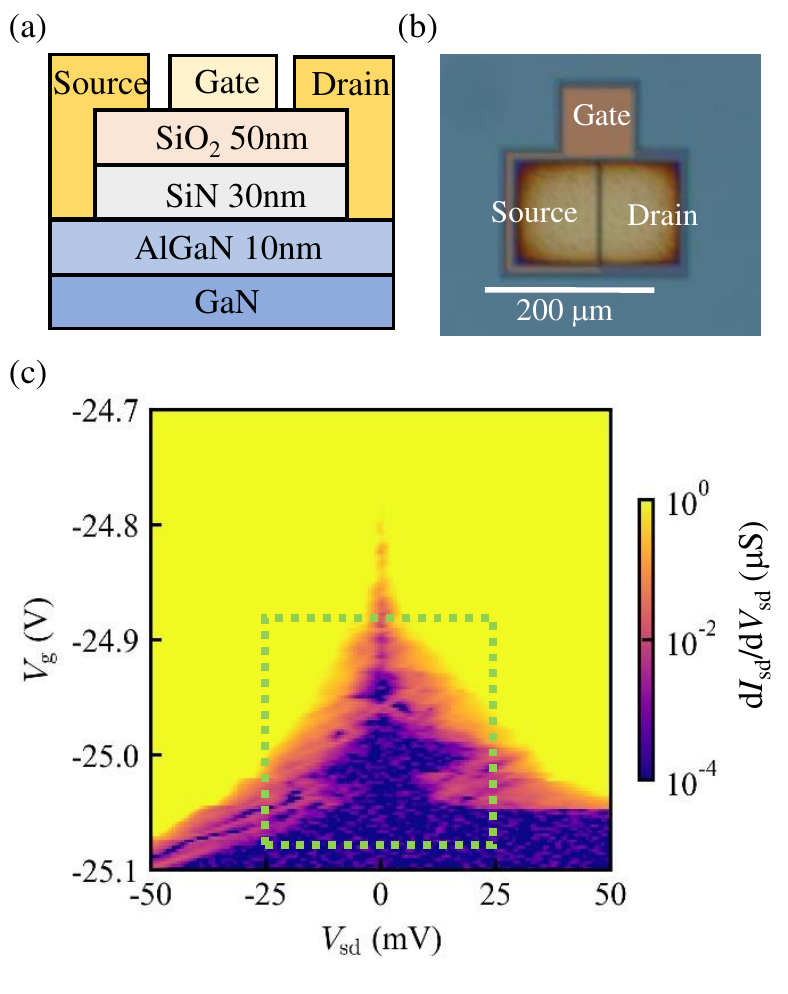}
  \caption{(a) Schematic of the layer structure of the device.
  (b) Optical image of the device.
  (c) Measured ${\rm d}I_{{\rm sd}}/{{\rm d}V_{{\rm sd}}}$ as a function of  $V_{\rm{sd}}$ and $V_{\rm{g}}$ near the pinch-off voltage of the FET device with 1.4~$\mu $m gate length at 2.3~K 
  }
  \label{device}
\end{center}
\end{figure}

The layer structure of the device is shown in Fig.~\ref{device}(a).
GaN and AlGaN layers are stacked on a Si substrate using chemical vapor deposition and the two-dimensional electron gas is formed at the interface between the GaN and AlGaN layers.
The typical values of electron density and mobility determined by a Hall measurement are 6.7$\times $10$^{12}$~cm$^{-2}$ and 1670~cm$^{2}$V$^{-1}$s$^{-1}$, respectively.
The source and drain electrodes are prepared using Ti and Al.
SiN and SiO$_2$ are used as gate insulators and TiN or Ti/Au is deposited on top of them as a gate electrode.
The gate length is 1.4, 1.0, 0.6, 0.4, 0.2 and 0.05~$\mu $m.
The gate width is 150~$\mu $m in all devices.
Figure~\ref{device}(b) shows the optical image of the device. 

The source-drain current $I_{\rm{sd}}$ are measured as a function of the source-drain voltage $V_{\rm{sd}}$ and the gate voltage $V_{\rm{g}}$.
The devices are cooled down to 2.3~K or 0.5~K using a helium depressurization refrigerator to measure the electronic transport at low temperature. 
Figure~\ref{device}(c) shows a typical result of the measurement near the pinch-off voltage of the FET device with 1.4~$\mu $m gate length at 2.3~K.
The differential conductance ${\rm d}I_{{\rm sd}}/{{\rm d}V_{{\rm sd}}}$ are plotted as a function of $V_{\rm{sd}}$ and $V_{\rm{g}}$.
The current changes non-monotonically with $V_{\rm{sd}}$ and $V_{\rm{g}}$ and Coulomb diamonds due to formation of quantum dots in the conduction channel are observed.
This formation is induced by disordered potential by impurities or defects, which creates potential minima confining electrons near the depletion condition of the two-dimensional electron gas\cite{2020OtsukaSciRep}.

\begin{figure}
\begin{center}
  \includegraphics{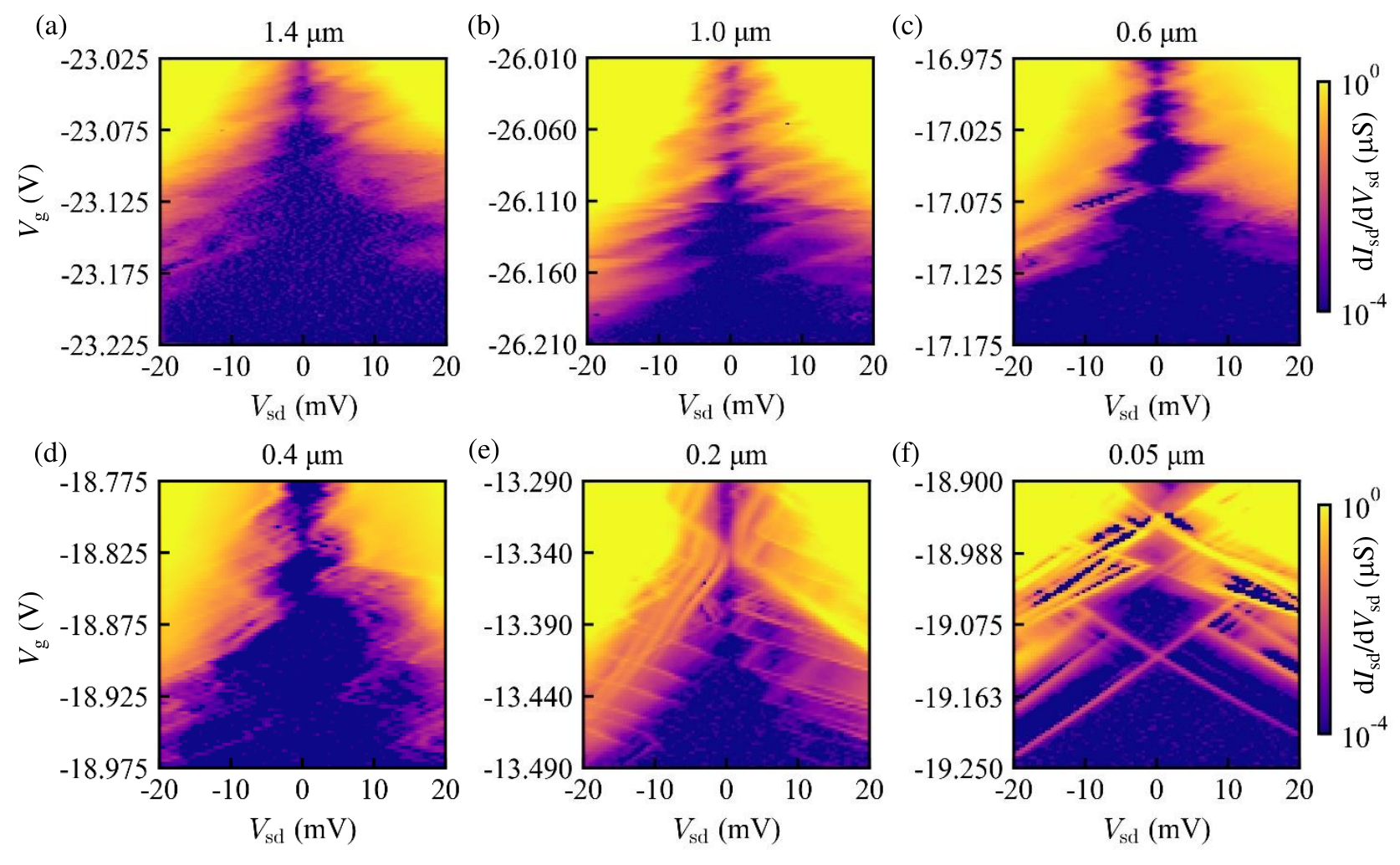}
  \caption{(a)-(f) Observed ${\rm d}I_{{\rm sd}}/{{\rm d}V_{{\rm sd}}}$ as a function of  $V_{\rm{sd}}$ and $V_{\rm{g}}$ in the devices with 1.4, 1.0, 0.6, 0.4, 0.2 and 0.05~$\mu $m gate length, respectively.
  }
  \label{Length}
\end{center}
\end{figure}

Next, we investigate the channel length dependence of the formation of quantum dots.
Figure~\ref{Length}(a)-(f) show results of the measurements in the devices with 1.4, 1.0, 0.6, 0.4, 0.2 and 0.05~$\mu $m gate length, respectively. 
We plot the results with areas in which the first Coulomb diamonds appear from the pinched-off states.
The size of the area is fixed at 0.2~V in $V_{\rm{g}}$ and -20 to +20 ~mV in $V_{\rm{sd}}$, except for the 0.05~$\mu $m gate length device. For the 0.05~$\mu $m gate length device, the size of the area is 0.35~V in $V_{\rm{g}}$.
The measurement temperature is 2.3~K (0.5 K) for (a) and (b) ((c), (d), (e) and (f)). The temperature difference does not affect the main features of the diamond although the fine structures are blurred a little bit in (a) and (b).

In the case of the 1.4~$\mu $m gate length device, this corresponds to the area enclosed by the dashed lines in Fig.~\ref{device}(c).
With the decrease of the channel length, the overlap of the multiple Coulomb diamonds at zero bias tends to decrease and the shape of the diamonds becomes clearer.
Especially in the 0.05~$\mu $m gate length device, a series of Coulomb diamonds and a completely closed Coulomb diamond are observed.
These results imply that the relatively smaller number of quantum dots are formed in devices with shorter gate length and a few quantum dots are formed in the 0.05~$\mu $m gate length device.

This channel length dependence can be explained as the following.
In long gate length devices, the numbers of the potential minima under the gate are large and multiple quantum dots are formed.
This induces a series coupling of the quantum dots and overlapping of Coulomnb diamonds.
With decreasing the gate length, we can reduce the numbers of the potential minima under the gate and the formed quantum dots.
Note that about the width, the current contributes in parallel and the minimum potential will determine the current near the pinch-off condition of the device.
Because we observe the change of the formation of the quantum dots by changing the gate length around by 100~nm, we can estimate the potential minima are distributed on a scale of several tenths to hundreds of nm.
This might be induced by interface states at the SiN/AlGaN interface\cite{2012YeluriJAP} or carbon dopants in the GaN layer\cite{2001TangAPL}.


\begin{figure}
\begin{center}
  \includegraphics{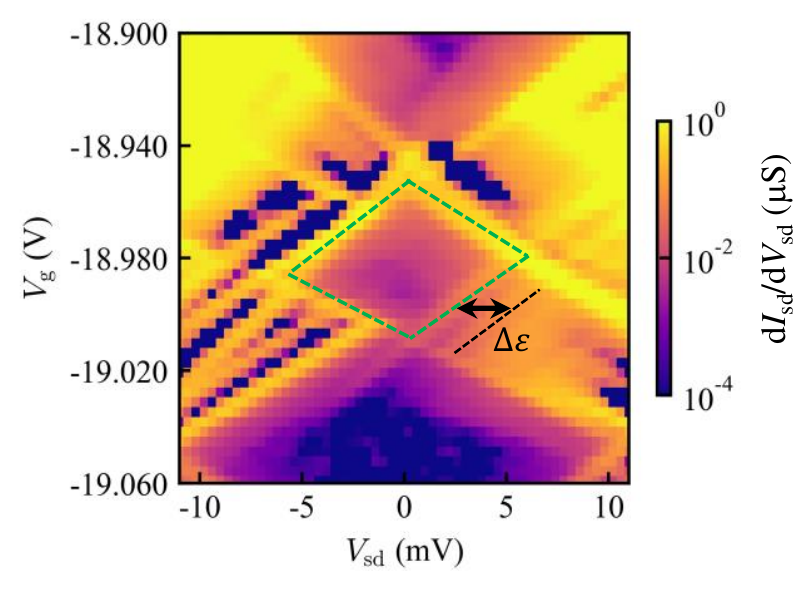}
  \caption{Closed up image of the Coulomb diamonds observed in the device with 0.05~$\mu $m gate length.
  }
  \label{calculation}
\end{center}
\end{figure}

Figure~\ref{calculation} shows the magnified image of the Coulomb diamonds observed in the device with 0.05~m gate length. From Coulomb diamond in $V_{\rm{sd}}$ direction, the charging energy of the quantum dot is evaluated as $E_{\rm c}\simeq$5.6~meV and the total capacitance as $C\simeq $29~aF.
From the spacing of the Coulomb peaks in the direction of $V_{\rm{g}}$ at zero bias, the capacitance between the quantum dot and the gate electrode is obtained as $C_{\rm{g}}\simeq$2.7~aF.
The slope of the edge of the Coulomb diamond can be expressed as $C_{\rm{g}}/C_{\rm{s}}$ and $C_{\rm{g}}/(C_{\rm{d}}+C_{\rm{g}})$.
The capacitances between the quantum dot and the source electrode and the drain electrode are evaluated as $C_{\rm{s}}\simeq $15~aF and $C_{\rm{d}}\simeq $12~aF, respectively.
The lever arm for converting the gate voltage to the electrostatic potential in the quantum dot becomes $\alpha = C_{\mathrm{g}}/C \simeq $0.09.
By assuming that the shape of the quantum dot is a disk, the size of the quantum dot is estimated from the value of $C_{\rm{g}}$ as 15~nm (see Supplementary Material).

We can also observe excited-state lines in Fig.~\ref{calculation}, which are parallel to the edge of the diamond.
From the distance between the edge of the diamond and the excitation line, the discrete energy level spacing of $\Delta\varepsilon = $2.9~meV is obtained.
Since the measurement temperature of 0.5~K is corresponding to an energy of 0.05~meV, it is reasonable to observe the discrete level spacing. Assuming the confinement potential as a harmonic oscillator, the size of the confinement is evaluated as 22~nm.
This value is consistent with the size evaluated from the capacitance between the quantum dot and the gate electrode. 
Note that the larger value evaluated from the excited state might be reflecting the asymmetry of the confinement potential in which lowers the orbital energy in the loose confinement direction.

\begin{figure}
\begin{center}
  \includegraphics{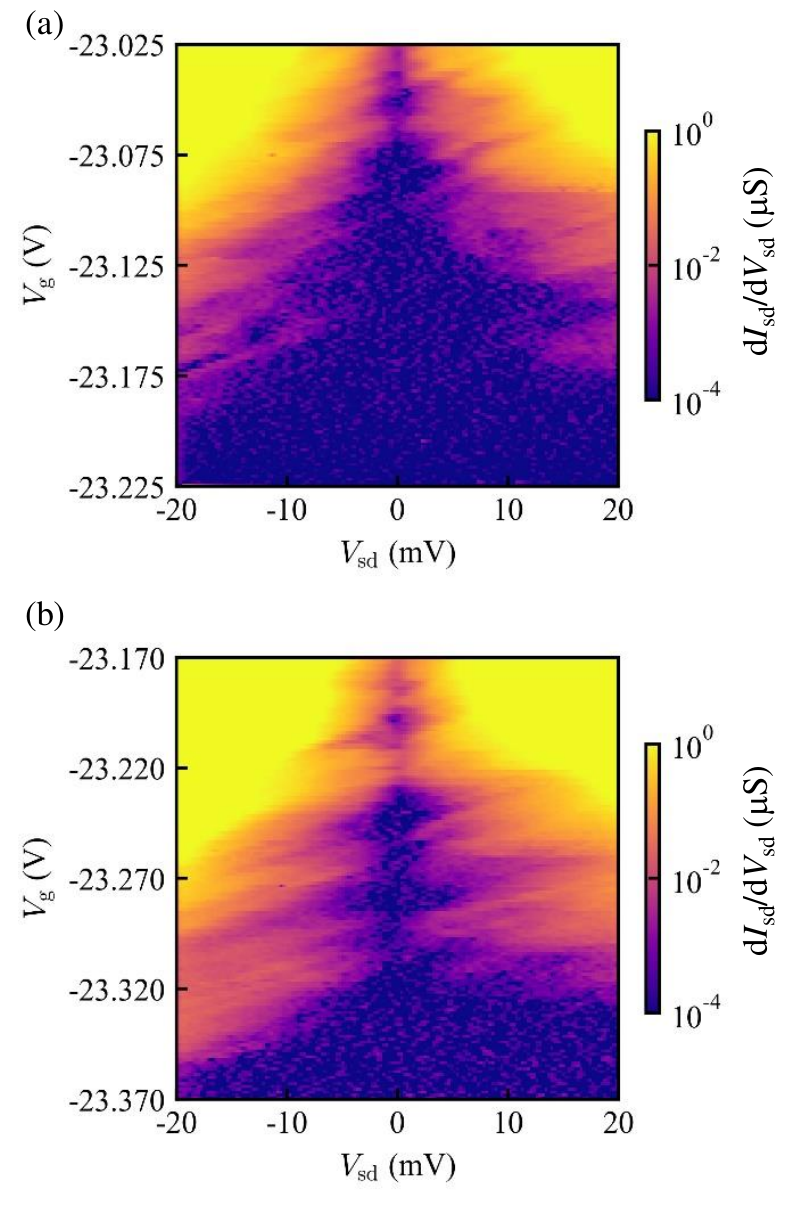}
  \caption{(a), (b) Coulomb diamonds measured before (a) and after the thermal cycle (b).
  }
  \label{Thermal}
\end{center}
\end{figure}

Next, we investigate the dependence of the quantum dot formation on a thermal cycle.
Because the quantum dots are induced by the disordered potential by impurities or defects, the spatial distribution will be modified by the change of the charge states of those by thermal excitation like in the cases of universal conductance fluctuations\cite{1987LeePRB, 1987ThorntonPRB}.
After the first measurement at 2.3~K, the sample is heated up to 300~K by using a heater and then cooled down to 2.3~K again.
To avoid the effect of photo-excitation, the sample is kept in dark condition without opening the refrigerator's cover in the thermal cycle.
The measured results of Coulomb diamonds in the sample with the channel length of 1.4~$\mu $m before and after thermal cycle are shown in Fig.~\ref{Thermal}(a) and (b), respectively.
Figure~\ref{Thermal}(a) is the same to Fig.~\ref{Length}(a).
The the Coulomb diamonds becomes different by the thermal cycle.
This is due to the change of the configuration of quantum dots.
Redistribution of the potential minima occurs by the thermal cycle.
Charge trapping sites which can be activated by the thermal energy at room temperature 26~meV contributes the redistribution.

\begin{figure}
\begin{center}
  \includegraphics{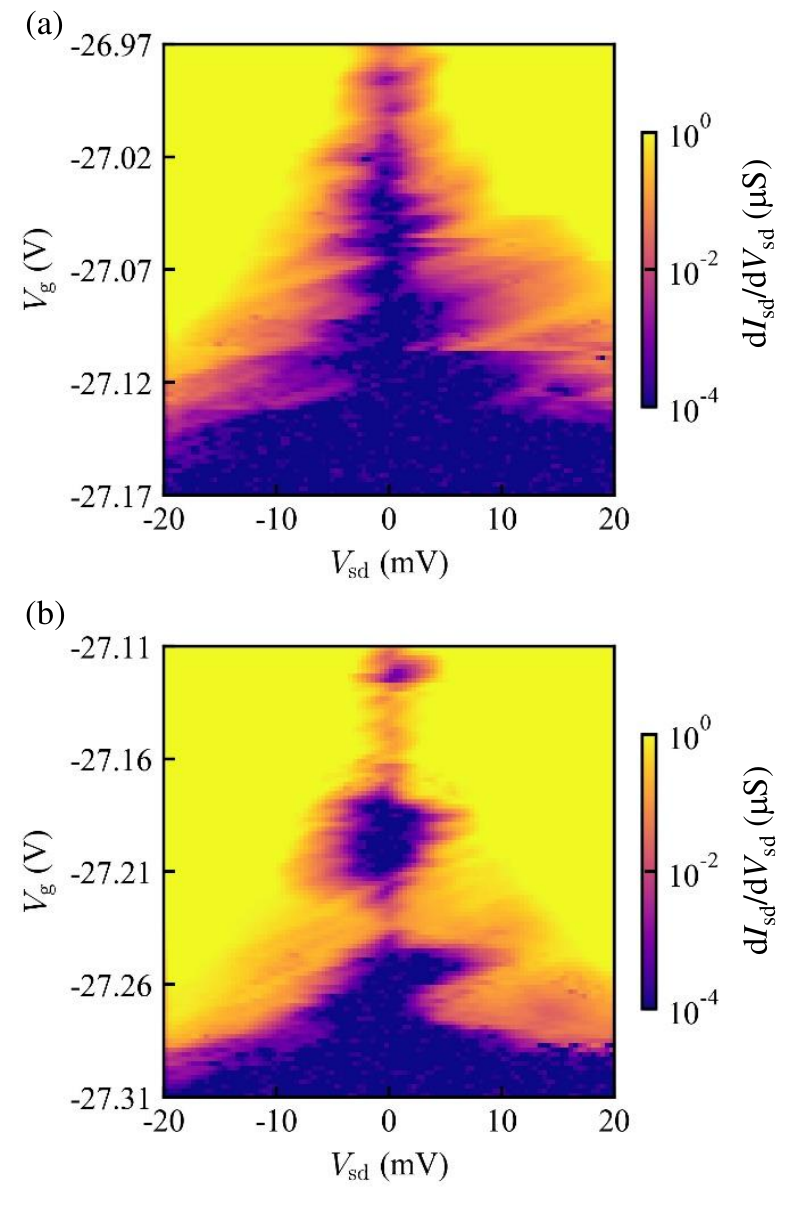}
  \caption{(a), (b) Coulomb diamonds measured before (a) and after (b) LED illumination.
  }
  \label{Light}
\end{center}
\end{figure}

Finally, we measured the effect of light illumination on the formation of the quantum dots.
The redistribution of the potential minima will be induced also by light.
After the first measurement at 2.3~K, a red LED attached near the sample is turned on for 180~seconds and then turned off with keeping the low temperature.
During this illumination, $V_{\rm{sd}}$ and $V_{\rm{g}}$ of the sample are fixed.
Figure~\ref{Light} (a) and (b) shows Coulomb diamonds measured in another sample with the channel length of 0.6~$\mu $m before and after LED illumination, respectively.
The pinch-off voltage shifts about -0.2 V due to the persistent photo-current\cite{1990FletchierPRB, 1997LiJApplPhys} and different Coulomb diamonds are observed.
Redistribution of the potential minima occurs by the illumination of light.
In this case, the trapping sites in SiN or AlGaN\cite{Nagarajan2020} might also contribute the reconfiguration of the quantum dots.

In conclusion, we investigate channel length dependence of the formation of quantum dots in GaN/AlGaN FETs.
The number of quantum dots formed in conduction channel decreases with the decrease of the channel length.
From the results, the distibution of the potential minima are estimated on a scale of several tenths to hundreds of nm.
In the case of 0.05~$\mu $m length device, a few quantum dots are formed and we evaluate quantum dot's parameters and the disordered potential.
We also measure the effects of a thermal cycle and light illumination on the formation of quantum dots.
These results are important in evaluation of the disordered potential in FET channels and development of quantum dot devices utilizing GaN/AlGaN.

\section{Acknowledgements}
The authors thank Alka Sharma and RIEC Fundamental Technology Center for fruitful discussions and technical supports.
Part of this work is supported by
Rohm Collaboration Project, 
MEXT Leading Initiative for Excellent Young Researchers, 
Grants-in-Aid for Scientific Research (20H00237, 21K18592), 
Telecom Advanced Technology Research Support Center Research Grant, 
Izumi Science and Technology Foundation Research Grant, 
Fujikura Foundation Research Grant, 
Hattori Hokokai Foudation Research Grant, 
and Tanigawa Foundation Research Grant.

K. M. and T. A. contributed equally to this work.

\end{document}